\documentclass[runningheads]{svmult}

\usepackage{makeidx}   
\usepackage{graphicx}  
\usepackage{subeqnar}  
\usepackage{multicol}  
\usepackage{physprbb}  
\makeindex             

\begin{document}
\title*{Pre-main-sequence Lithium Depletion}
\toctitle{Pre-main-sequence Lithium Depletion}

%
%
\titlerunning{PMS Li Depletion}
%
\author{R. D. Jeffries}

\authorrunning{R. D. Jeffries}
%
%
\institute{Astrophysics Group, School of Chemistry and Physics, Keele
  University, Staffordshire, ST5 5BG, UK}

\maketitle              

\begin{abstract}
In this review I briefly discuss the theory of pre-main-sequence (PMS)
Li depletion in low-mass ($0.075<M<1.2\,M_{\odot}$) stars and highlight
those uncertain parameters which lead to substantial differences in 
model predictions. I then summarise observations of PMS stars in very
young open clusters, clusters that have just reached the ZAMS and
briefly highlight recent developments in the observation of Li in very
low-mass PMS stars.
\end{abstract}

\section{Introduction}
During pre-main-sequence (PMS) evolution,
Li is burned at relatively low temperatures (2.5--3.0$\times 10^{6}$\,K) and,
in low-mass stars ($<1.2\,M_{\odot}$), convective mixing can rapidly bring Li-depleted
material to the photosphere. For this reason, photospheric Li
abundance measurements provide one of the few methods of probing
stellar interiors and are a sensitive test of PMS evolutionary
models. Understanding PMS Li depletion also offers a route to
estimating the ages of young stars and of course is a pre-requisite for
quantifying any subsequent main-sequence Li depletion
(see Randich 2005, these proceedings).

\section{Models of PMS Li depletion}

\subsection{Very low-mass stars}

\begin{figure}[b]
\begin{center}
\includegraphics[width=1.0\textwidth]{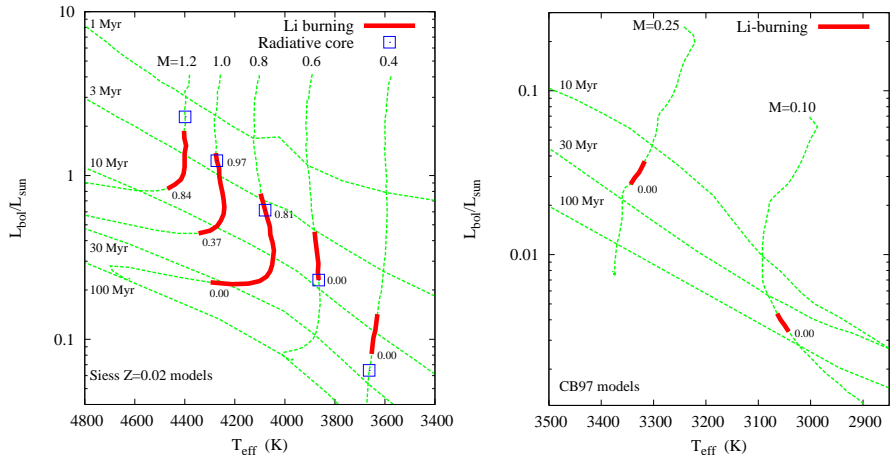}
\end{center}
\caption[]{Evolutionary tracks (labelled in $M_{\odot}$) and isochrones
  (in Myr) for low-mass stars taken from two models
  \cite{chabrier97,siess00}. 
  The epochs of
  photospheric Li depletion (and hence Li-burning in the core of a
  fully convective star or at the convection zone base otherwise) and
  the development of a radiative core are indicated. The numbers to the
  right of the tracks 
  indicate the fraction of photospheric Li remaining at the point where
  the radiative core develops and at the end of Li burning.}
\label{massplot}
\end{figure}

PMS stars with $M<0.35\,M_{\odot}$ have a simple structure -- they are
fully convective balls of gas all the way to the ZAMS. As the star
contracts along its Hayashi track the core heats up, but the
temperature gradient stays very close to adiabatic except in the
surface layers. Li begins to burn in $p,\alpha$ reactions when the core
temperature, $T_{c}$ reaches $\simeq 3\times10^{6}$\,K and, because the reaction
is so temperature sensitive ($\propto T_{c}^{16-19}$ at typical PMS
densities) and convective mixing so very rapid, all the Li is burned in
a small fraction of the Kelvin-Helmholtz timescale (see Fig.~\ref{massplot}).

The age at which Li depletion occurs increases with decreasing mass
(and Li-burning temperatures are never reached for
$M<0.06\,M_{\odot}$). As luminosity, $L \propto M^{2}$ for PMS stars,
the luminosity at which complete Li depletion takes place is therefore
a sensitive function of age between about 10 and 200\,Myr~\cite{bildsten97}. This
relationship depends little on ingredients of the PMS models such as
the treatments of convection and interior radiative opacities because
the stars are fully convective. The extreme temperature dependence
means nuclear physics uncertainties play little role, and there is only
a small dependence on the kind of atmosphere assumed as a boundary
condition, or the adopted equation of state. Indeed, whilst the chosen
form of the atmosphere (grey or non-grey) changes the $T_{\rm eff}$ at
which Li is burned, it hardly affects the luminosity. Ages determined from the luminosity
at the "Li depletion boundary" (LDB) vary by only 10 per cent between
different models and even analytical treatments~\cite{burke04}.

\subsection{Higher mass stars}

Li depletion is {\it much} more complex in higher mass stars. They have
lower central densities and as $T_{c}$ rises during PMS contraction,
the opacity falls sufficiently for the temperature gradient to become
sub-adiabatic. A radiative core forms which pushes outward to include a
rapidly increasing fraction of the stellar mass. For $M<1\,M_{\odot}$
there is small window of opportunity to burn some Li before the
radiative core develops (at $\simeq 2$\,Myr for 1\,$M_{\odot}$). For
$M<0.6\,M_{\odot}$ all the Li is burned in this way (see
Fig.~\ref{massplot}).  For higher mass stars the radiative core
develops before Li burning is complete and the temperature at the base
of the convective envelope, $T_{bcz}$, decreases.  In the absence of
convective mixing, Li-depleted material cannot get to the photosphere,
so once $T_{bcz}$ drops below the Li-burning threshold, photospheric
Li-depletion ceases.  Photospheric Li depletion begins at about 2\,Myr in a
1\,$M_{\odot}$ star and should terminate at about 15\,Myr. This window
shifts towards older ages in lower mass stars. However, the overall
amount of Li-depletion is extremely sensitive to mass (and other model
parameters -- see below). There should be relatively little depletion
in solar mass stars compared with lower-mass stars (see Fig.~\ref{pleiadesplot}).

The exact amount of Li depletion expected is exquisitely dependent on a
number of model details. The reason is that whilst Li depletion is
occurring, even with a radiative core, the overall temperature gradient
in the stars is still very close to adiabatic (see Fig.~2 in
\cite{piau02}). It takes only a small perturbation to this gradient to
change the time at which the radiative core develops, the position of
the convection zone base and hence $T_{bcz}$. As a result large changes
in Li depletion predictions can result from relatively minor
perturbations in model parameters. Similarly, because photospheric
Li-depletion arises from rapid Li burning in a very thin region above
the convection zone base, a model grid with temporal and spatial
resolution merely sufficient to model the structure of the star may be
an order of magnitude to coarse to accurately predict Li
depletion~\cite{piau02}.

Convective efficiency is a crucial model parameter. If convection is
efficient then $T_{bcz}$ is higher (at a given mass) and hence
stays above the Li-burning threshold for longer, resulting in much more
photospheric Li depletion \cite{dantona03}. A typical approach to modelling convection
is to use mixing length theory with the mixing length set by requiring a
model to reproduce the solar structure (revealed by helioseismology) at
the age of the Sun. It is not clear that this approach is valid. The
mixing length may vary with time, depending on evolutionary stage, surface gravity or
effective temperature. Adopting alternate convection theories, such as the full
spectrum of turbulence models which have more efficient convection in
the deep layers, results in orders of magnitude more PMS Li depletion at the
same mass (see \cite{dam97} and Fig.~\ref{pleiadesplot}).

Opacity effects are also important. This can refer to differences in
the treatment of interior opacities or to the effects of uncertain
stellar compositions on the opacities. An increase in opacity makes
temperature gradients larger, keeps the star convective for longer,
raises $T_{bcz}$ once the radiative core develops and so leads to
enhanced Li depletion. Opacity is increased by an increase in overall
metallicity or a decrease in the Helium abundance. Changes of only 0.1
dex in metallicity can lead to an order of magnitude change in Li
depletion (e.g. see Fig.~2 of~\cite{ventura98}).

Other factors, such as the adopted equation of state or chosen
treatment of the atmospheric boundary conditions have some effect on
Li-depletion predictions, but are much less significant.

\section{Observations}

There have been more measurements of Li in stars than any other
chemical element. The vast majority have been derived from high
resolution spectra of the strong Li\,{\sc i}\,6708\AA\ resonance
doublet. Only a fraction of the observational material can be reviewed
here. The reader is referred to some other reviews for a more complete
picture~\cite{jeffries00,pallavicini00}.

\subsection{The initial Li abundance}
Theory doesn't tell us what initial Li a star has, only what
depletion it suffers. An accurate estimate of the initial Li abundance
is therefore a pre-requisite before observations and models can be
compared. The Sun is a unique exception, where we know the present abundance,
$A($Li$)=1.1\pm0.1$ (where $A$(Li)$=\log[N$(Li)$/N$(H)$]+12$) and the
initial abundance of $A$(Li)$=3.34$ is obtained from meteorites.
For recently born stars, the initial Li abundance is estimated from
photospheric measurements in young T-Tauri stars, or from the hotter F stars of
slightly older clusters, where {\it theory suggests} that no Li
depletion can yet have taken place. Results vary from
$3.0<A$(Li)\,$<3.4$, somewhat dependent on assumed atmospheres, NLTE
corrections and $T_{\rm eff}$ scales~\cite{martin94,sod99}. 
It is of course quite possible
that the initial Li, like Fe abundances in the solar neighbourhood, shows some
cosmic scatter. Present observations certainly cannot rule this out, leading
to about a $\pm 0.2$\,dex systematic uncertainty when comparing
observations with Li depletion predictions.

\subsection{ZAMS clusters}

\begin{figure}[b]
\begin{center}
\includegraphics[width=0.65\textwidth]{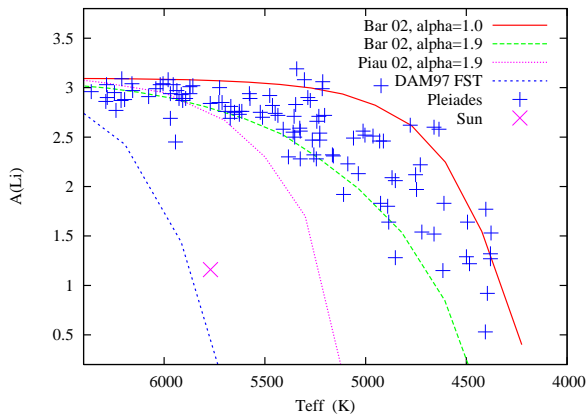}
\end{center}
\caption[]{Measured Li abundances for the Sun and the
  Pleiades~\cite{sod93} compared with a variety of
  models~\cite{bar02,dam97,piau02}. The majority of the differences are
  due to the convective efficiency used by the models}
\label{pleiadesplot}
\end{figure}

Clusters that are old enough for stars to have reached the ZAMS
empirically show us the results of PMS Li depletion. The canonical
dataset is that for the Pleiades (Fig.~\ref{pleiadesplot}, \cite{sod93}). 
With an age of 120\,Myr, all stars
with $M>0.5\,M_{\odot}$ have reached the ZAMS. Assuming an
initial $A$(Li) of 3.2, then there seems to have been little PMS Li
depletion among F-stars, $\leq 0.2$\,dex in G stars and then a
strongly increasing level of Li depletion with decreasing mass. There
is also evidence for a {\it scatter} in Li abundances that develops for
$T_{\rm eff}<5300$\,K and probably continues to $T_{\rm eff}\simeq
4000$\,K, where Li becomes undetectable~\cite{garcia94}. Similar results are now
available for a number of clusters with ages 50-200\,Myr
(e.g.~\cite{barrado01,hunsch04,jeffries98,martin97,randichic2391}).

The difference in the Li abundances in the G-stars of the Pleiades and
the Sun, combined with the probable similarities in their overall
chemical composition tell us that PMS Li depletion cannot be the whole
story. Another mechanism, additional to convective mixing, must be
responsible for Li depletion whilst solar-type stars are on the
main-sequence. Recent PMS models that have their convective treatments
tuned to match the structure of the Sun reproduce the mass dependence
of Li depletion, but deplete too much Li compared
with the Pleiades, and can even explain the solar $A$(Li) in
the case of full spectrum turbulence models~\cite{dam97}. The
over-depletion with respect to the Pleiades gets worse at lower
masses. Better fits to the Pleiades data are achieved with PMS models that
feature relatively inefficient convection with smaller mixing lengths.

\subsection{An Li abundance scatter?}
The apparent scatter among Li abundances in K-type and lower mass stars
of the Pleiades and other young clusters is intriguing. It is either
telling us something about the physics of mixing and Li-burning inside
PMS stars or it is telling us something about the atmospheres of these
stars such that we cannot properly estimate their Li abundances.  Clues
include: the $T_{\rm eff}$ at which the scatter develops, which
coincides with those stars that did most of their Li depletion in a
fully convective state; and the strong correlation between apparent Li
abundance and rotation rate for the K-type stars, such that fast
rotators appear to have high $A$(Li), whereas slower rotators can have
either higher or lower $A$(Li) than average. This correlation may be
weaker or absent in the lower mass stars~\cite{garcia94,jones96}.  
Efforts to
understand the apparent Li abundance scatter divide into those that
propose a physical mechanism for the Li abundance scatter (i.e. that
assume the scatter is real) and those that assume the scatter is not
real and instead suggest that the strength of the
Li\,{\sc i}\,6708\AA\ feature does not reliably yield true Li abundances.

{\it Rotationally Driven Mixing:} Non-convective mixing can take place in
radiative regions, driven by angular momentum loss (AML), and causes
additional Li depletion. Fast rotating ZAMS stars have suffered little
AML and so would have the highest Li abundances. Slow rotators may have
undergone little AML (if they started out with less angular momentum),
or lots (if they remained magnetically coupled to a circumstellar disc
for an extended period) and so could have a range of Li abundances.
Problems with this persuasive picture are that {\it additional} PMS Li
depletion is predicted, widening the disagreement between solar-tuned
models and ZAMS clusters and that very little scatter is actually
produced in theoretical models even with a realistic range of initial
angular momenta~\cite{pinsonneault99}.

{\it Structural Effects of Rotation:} Rapid rotation in a fully
convective star decreases the core temperature, but actually increases
$T_{bcz}$ once a radiative core has developed. The net effect on Li
depletion seems to be rather small and cannot explain the dispersion of
Li abundances seen among the slow rotating ZAMS stars~\cite{mendes99}.

{\it Composition Variations:} Li depletion is sensitive to interior
opacities, which themselves depend on the stellar composition. Small
star-to-star variations might cause an Li abundance scatter, which
would grow towards lower masses. However, current limits on metallicity
variations in the Pleiades (and other clusters) 
seem too small for this to be the dominant explanation of
any scatter~\cite{wilden02}. In addition, the correlation of
Li-depletion with rotation is unexplained.

{\it Accretion:} Li abundances can be altered in two ways by
accretion. During PMS Li depletion the additional mass will lead to
increased Li depletion at a given $T_{\rm eff}$ when the star reaches
the ZAMS~\cite{piau02}. If accretion occurs after Li-burning has ceased then the
convective zone is enriched with Li. Too much accretion is required to be
compatible with observations of disks around PMS stars unless the
accreted material is H/He-deficient. But then accretion of sufficient
H/He depleted material to explain the Li abundance scatter would also
lead to (for instance) Fe abundance anomalies of order 0.2-0.3\,dex --
much higher than allowed by current observational constraints~\cite{wilden02}.

{\it Magnetic Fields:} Low-mass PMS stars are known to be magnetically
active. B-fields in the convection zone can provide additional
support, raise the adiabatic temperature gradient, hasten the onset of
a radiative core and hence decrease Li depletion. Magnetic activity
{\it may} be correlated with rotation in PMS stars at the critical ages
of 2-20\,Myr but this remains to be established. Basic models including
B-fields in the convection zone have now been
developed~\cite{dantona00,ventura98}, suggesting
this mechanism could inhibit Li depletion by orders of magnitude!

{\it Atmospheric effects:} The atmospheres of PMS stars are doubtless
more complicated than the 1-d, homogeneous models usually used to
estimate their Li abundances. Starspots and plages complicate the
interpretation and could lead to a scatter in the strength of Li\,{\sc
  i} spectral features at a given abundance~\cite{barrado99,ford02}. 
The 6708\AA\ line is also
formed high in the atmosphere and is susceptible to NLTE effects and
possible overionisation from an overlying chromosphere~\cite{stuik97}. It is telling
that the analogous K\,{\sc i} resonance line mimics the behaviour of
the Li\,{\sc i} line, despite there being no possibility of significant
K abundance variations~\cite{king00,randich01,sod93}. 
Varying activity levels could at least be
responsible for some of the {\it apparent} Li abundance
scatter. Arguing against this are that very little time variability is
seen in the strength of the Li\,{\sc i}\,6708\AA\ line, despite
magnetic/chromospheric activity being quite variable in cool ZAMS
stars~\cite{jeffries99}. In addition, measurements of the 
weak Li\,{\sc i}\,6104\AA\ feature,
which is probably less susceptible to details of the model atmosphere,
have implied a scatter in Li abundances at least as large as that
derived from the resonance line~\cite{ford02}.

\subsection{The metallicity dependence of PMS Li depletion}

PMS Li depletion is supposed to be very sensitive to overall
metallicity. Groups of ZAMS clusters with similar ages but differing
metallicities can be used to test this prediction. The results are
surprising. Metallicity variations of 0.1-0.2\,dex appear to make no
difference to PMS Li depletion~\cite{barrado01,jeffriesblanco99}. 
An explanation might be that whilst
[Fe/H] (what is usually measured as a proxy for metallicity) varies,
other elements which are important for interior opacities, especially
O, Si, Mg, might vary in the opposite direction to compensate. Quite
small differences of 0.1--0.2\,dex 
in [O/Fe] would be required~\cite{piau02}, but these differences
are still uncomfortably high compared with the spread in [O/Fe]
measured for field dwarfs~\cite{reddy03}. In addition, it would require a cosmic
conspiracy of some proportions to ensure that the half dozen ZAMS
clusters investigated so far, all had similar interior opacities.
Careful and consistent multi-element abundance determinations are
required for these clusters to definitively address the issue.

An interesting aside to this discussion concerns the composition mix
assumed in the theoretical models. Recent measurements have suggested
that the solar O abundance might be 0.2\,dex lower than previously
believed~\cite{asplund04}. A change of this size in the model compositions could lead to
significantly less PMS Li depletion among solar-type stars, reducing
the discrepancy between the Li depletion predicted by solar-tuned
convective models and the ZAMS cluster data.

\subsection{Very low-mass stars}
Whilst problems remain in the modelling and interpretation of
$0.6<M<1.2\,M_{\odot}$ stars, the situation is more favourable in lower
mass objects that are always fully convective.  In agreement with
theory, observations of four young clusters (Pleiades, Alpha Per, IC
2391 and NGC 2547) have now found the sharply defined LDB, where the
original undepleted Li abundance is seen in the coolest objects and
which marks the age-dependent point at which cores are still too cool
to burn Li~\cite{barrado04,jeffries05,stauffer98,stauffer99}. 
Because the LDB is a model-insensitive chronometer, these
LDB ages can be used to test the physics which goes into isochronal
ages determined from higher-mass stars. The conclusions are that
LDB ages are 50 per cent older than nuclear turn-off ages without
convective core overshoot, but in reasonable agreement with isochronal
ages defined by the descent of low-mass stars to the ZAMS.

\section{Summary}

The study of PMS Li depletion divides into two regimes. For very low
mass stars $0.075<M<0.35\,M_{\odot}$, the few extant observations are
fully in agreement with available theoretical predictions. Furthermore
there is little variation in the predictions of different models and
little dependence on uncertain physical processes or
parameters. However, models of PMS Li depletion for higher mass stars
($0.35<M<1.2\,M_{\odot}$ in which a radiative core develops, make
wildly varying (by orders of magnitude) quantitative predictions for Li
depletion. None of these models satisfactorily explain all aspects of
the data, particularly the presence of an apparent scatter in Li
abundances at the end of the PMS phase and the lack of any sensitivity
of Li-depletion to stellar metallicity. The model dependence does at
least give hope that some aspects of PMS evolution may ultimately be
tightly constrained by Li abundance measurements.  As an example the
current data-model comparisons suggest that PMS convective efficiency
is lower than suggested by tuning models to produce the Sun,
particularly among cooler stars.

%


\begin{thebibliography}{8.}
\addcontentsline{toc}{section}{References}

\bibitem{asplund04} M. Asplund, N. Grevesse, A. Jacques Sauval: `The
  solar chemical composition'. In: \emph{Cosmic abundances as records
  of stellar evolution and nucleosynthesis}, ed. by F.N. Bash,
  T.G. Barnes (ASP San Francisco 2005), in press 

\bibitem{barrado01} D. Barrado y Navascu\'es, C.P. Deliyannis,
  J.R. Stauffer: ApJ \textbf{549}, 452 (2001)

\bibitem{barrado99} D. Barrado y Navascu\'es, R.J. Garc\'ia-L\'opez,
  G. Severino, M.T. Gomez: A\&A \textbf{371}, 652 (2001)

\bibitem{barrado04} D. Barrado y Navascu\'es, J.R. Stauffer,
  R. Jayawardhana: ApJ \textbf{614}, 386 (2004)

\bibitem{bar02} I. Baraffe, G. Chabrier, F. Allard, P.H. Hauschildt:
  A\&A \textbf{382}, 563 (2002)

\bibitem{bildsten97} L. Bildsten, E.F. Brown, C.D. Matzner,
  G. Ushomirsky: ApJ \textbf{482}, 442 (1997)

\bibitem{burke04} C.J. Burke, M.H. Pinsonneault, A. Sills: ApJ
  \textbf{604}, 272 (2004)


\bibitem{chabrier97} G. Chabrier, I. Baraffe: A\&A \textbf{327}, 1039 (1997)

\bibitem{dam97} F. D'Antona, I. Mazzitelli:
  Mem. Soc. Astr. It. \textbf{68}, 807 (1997)

\bibitem{dantona03} F. D'Antona, J. Montalb\'{a}n: A\&A \textbf{412}
  213 (2003)

\bibitem{dantona00} F. D'Antona, P. Ventura, I. Mazzitelli: ApJ
  \textbf{543}, L77 (2000)
 
\bibitem{ford02} A. Ford, R.D. Jeffries, B. Smalley: A\&A \textbf{391}, 253
(2002)

\bibitem{garcia94} R.J. Garc\'ia-Lopez, R. Rebolo, E.L. Mart\'in: A\&A
  \textbf{282}, 518 (1994)

\bibitem{hunsch04} M. H\"unsch, S. Randich, M. Hempel, C. Weidner,
  J.H.M.M.~Schmitt: A\&A \textbf{418}, 539 (2004)

\bibitem{jeffries99} R.D. Jeffries: MNRAS \textbf{304}, 821 (1999)

\bibitem{jeffries00} R.D. Jeffries: `Lithium depletion in open
  clusters'. In: \emph{Stellar Clusters and Associations: Convection,
  Rotation, and Dynamos, ASP Proceedings Vol. 198}, ed. by R. Pallavicini, G. Micela, and
  S. Sciortino (ASP, San Francisco 2000) p.245

\bibitem{jeffriesblanco99} R.D. Jeffries, D.J. James: ApJ \textbf{511}, 218 (1999)

\bibitem{jeffries98} R.D. Jeffries, D.J. James, M.R. Thurston: MNRAS
  \textbf{300}, 550 (1998)

\bibitem{jeffries05} R.D. Jeffries, J.M. Oliveira: MNRAS, in press (2005)

\bibitem{jones96} B.F. Jones, M. Shetrone, D. Fischer, D.R.~Soderblom:
  AJ \textbf{112}, 186 (1996)

\bibitem{king00} J.R. King, A. Krishnamurthi, M.H. Pinsonneault: AJ
  \textbf{119}, 859 (2000)


\bibitem{martin97} E.L. Mart\'{i}n, D. Montes: A\&A \textbf{318}, 805
  (1997)

\bibitem{martin94} E.L. Mart\'{i}n, R. Rebolo, A. Magazz\`{u},
  Ya.V. Pavlenko: A\&A \textbf{282}, 503 (1994)

\bibitem{mendes99} L.T.S. Mendes, F. D'Antona, I.~Mazzitelli: A\&A
  \textbf{341}, 174 (1997)

\bibitem{pallavicini00} R. Pallavicini, S. Randich, J.R. Stauffer,
  S.C. Balachandran: `Lithium in young open clusters'. In: \emph{The
  Light Elements and their Evolution, Proc. IAU Symp. 198}, ed. by
  L. da Silva, R. de Medeiros, M Spite (2000) p.350

\bibitem{piau02} L. Piau, S. Turck-Chi\`{e}ze: ApJ \textbf{566}, 419 (2002)

\bibitem{pinsonneault99} M.H. Pinsonneault, T.P. Walker, G. Steigman,
  V.K.~Narayanan: ApJ \textbf{527}, 180 (1999)

\bibitem{randich01} S. Randich: A\&A \textbf{377}, 512 (2001)

\bibitem{randichic2391} S. Randich, R. Pallavicini, G. Meola,
  J.R. Stauffer, S.C. Balachandran: A\&A \textbf{372}, 862 (2001)

\bibitem{reddy03} B.E. Reddy, J. Tomkin, D.L. Lambert, C. Allende
  Prieto: MNRAS \textbf{340}, 304 (2003)

\bibitem{siess00} L. Siess, E. Dufour, M. Forestini: A\&A
  \textbf{358}, 593 (2000)

\bibitem{sod93} D.R. Soderblom, B.F. Jones, S. Balachandran,
  J.R.~Stauffer, D.K.~Duncan, S.B.~Fedele, J.D.~Hudon: AJ \textbf{106},
  1059 (1993)

\bibitem{sod99} D.R. Soderblom, J.R. King, L. Siess, B.F.~Jones,
  D.~Fischer: AJ \textbf{118}, 1301 (1999)

\bibitem{stauffer98} J.R. Stauffer, G. Schultz, J.D. Kirkpatrick: ApJ
  \textbf{499}, L199 (1998)

\bibitem{stauffer99} J.R. Stauffer et al.: ApJ \textbf{527}, 219 (1999)

\bibitem{stuik97} R. Stuik, J.H.M.J. Bruls, R.J.~Rutten: A\&A
  \textbf{322}, 911 (1997)

\bibitem{ventura98} P. Ventura, A. Zeppieri, I. Mazzitelli,
  F. D'Antona: A\&A \textbf{331}, 1011 (1998)

\bibitem{wilden02} B.S. Wilden, B.F. Jones, D.N.C. Lin, D.R.~Soderblom:
  AJ, \textbf{124}, 2799 (2002)


\end{thebibliography}
\end{document}